\documentclass[
  ,draft            
  ]
  {aipproc}

\layoutstyle{6x9}

\begin{document}

\title{Aperture Masking Interferometry for Subaru}

\classification{}
\keywords      {Technique:Interferometry, Adaptive Optics}

\author{Frantz Martinache}{address={Subaru Telescope}}
\author{Olivier Guyon}{address={Subaru Telescope}}
\author{Vincent Garrel}{address={Subaru Telescope}}

\begin{abstract}
Aperture Masking Interferometry used in combination with Adaptive
Optics, is a powerful technique that permits the detection of faint
companions at small angular separations. The precision calibration of
the data achieved with this observing mode indeed leads to reliable
results up to and beyond the formal diffraction limit, explaining why
it has, in just a few years, been ported on most major telescopes.
In this poster, we present its possible implementation on Subaru. We
also discuss how the opportunity offered by the planned Extreme-AO
upgrade to HiCIAO will push further the performance of this already
successful technique, offering Subaru a unique access to a very
exciting region of the "contrast-ratio - angular separation" parameter
space.
\end{abstract}

\maketitle


\begin{figure}
  \resizebox{\textwidth}{!}{\includegraphics{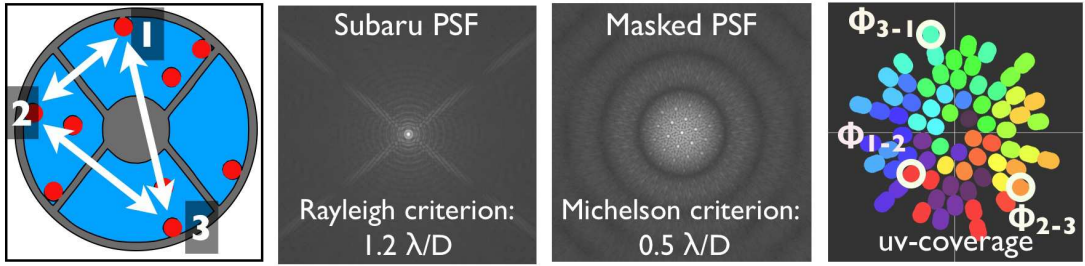}}
  \caption{
    Example of non-redundant aperture mask that accomodates the
    Subaru telescope pupil.}
  \label{fig:subaru_mask}
\end{figure}

Initially developed for radio astronomy \citep{1958MNRAS.118..276J},
non-redundant baseline interferometry is now used in IR and optical
bandpasses \citep{1986Natur.320..595B, 1998SPIE.3350..839T}. 
A modern description of NRM can be found in
\citet{2000PASP..112..555T, 2003RPPh...66..789M} and references
therein.
In brief, an N-hole pupil mask turns the extremely redundant
full aperture of a telescope into a simpler interferometric
array. Taking into account the geometry of the pupil (segmentation,
central obscuration and spider vanes) and the bandwidth, the
mask is designed such that each baseline (a vector linking the centers
of two holes) is unique: the mask is said non-redundant in the sense
that each spatial frequency is only sampled once
(Fig. \ref{fig:subaru_mask}).

In a sense, NRM is a form of \textit{imaging} with an unsual-looking
PSF that displays multiple sharp peaks, whereas a traditional
diffraction-limited PSFs is dominated by one only peak. The real
virtues of the NRM PSF however appear once the image is Fourier
transformed (e.g. Fig. \ref{fig:subaru_mask} right panel) as all
spatial frequencies sampled by the mask appear as well separated peaks
containing both amplitude and phase information.
The advantages of this approach are that:

\begin{itemize}\addtolength{\itemsep}{-0.1\baselineskip}
\item
The longest baseline provides a resolving power of $0.5\lambda/D$ to
be compared the traditionally accepted limit of $1.22 \lambda/D$ known
as the Rayleigh criterion.
\item
Unlike conventional PSF co-addition, the {\bf information}
extracted from the NRM data {\bf can be averaged} to reduce noise,
even in the presence of slowly-varying speckles.
\item
Non-redundancy ensures that the complex visibilities can be used to
form {\bf closure phases} \citep{1986Natur.320..595B}, an observable
that {\bf calibrates} wavefront residual errors as well as {\bf
  non-common path errors} between the science and sensing arms of the
instrument.
\item
The outer working angle (OWA) is set by the shortest baseline,
typically $4 \lambda/D$: the {\bf NRM} search space nicely
{\bf complements} that of {\bf coronagraphy}.
\end{itemize}

It may seem perverse to throw away most of a telescope's collecting
area, but the $10-20 \%$ mask transmission price paid for NRM
(comparable to most high performance coronagraphs' throughput)
purchases not only a straightforward $2.44\times$ gain in resolution
but also a dramatic increase in signal to noise ratio.
Today's ground-based NRM routinely achieves stability of 0.5 degrees
on the closure phase, hence passively stabilizing the phase at the
level of $\lambda/500$ to $\lambda/1000$, a performance that will only
be matched by the next generation of ExAO coronagraphic systems
\citep{2008ApJ...688..701S}.

\begin{figure}
  \resizebox{.65\textwidth}{!}{\includegraphics{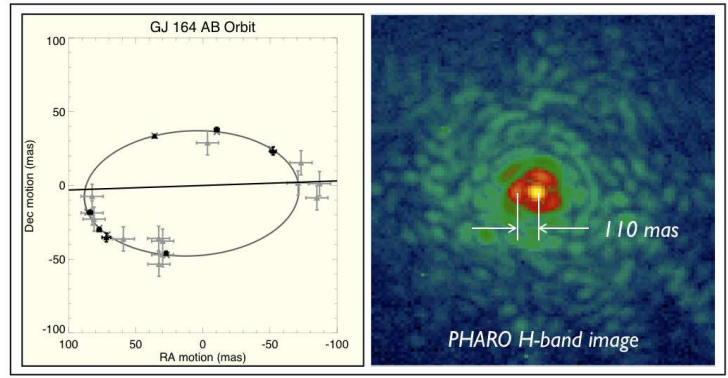}}
  \caption{
    Example of result obtained with the proposed technique with the
    instrument PHARO at the Cassegrain focus of the Palomar Hale
    telescope. The 5-meter aperture was able to resolve (black points)
    the M8 companion orbiting the star GJ 164 at separations between
    ranging from 40 to 80 mas, in H- and K-bands \citep{2009ApJ...695.1183M}.}
\end{figure}

\bibliography{martinache}

\end{document}